\documentclass[aps,pre,preprint,showpacs]{revtex4}

\usepackage{graphicx}

\begin{document}
\title{Closed model for granular compaction under weak tapping}
\author{J. Javier Brey}
\email[]{brey@us.es}
\author{A. Prados}
\email[]{prados@us.es}

\affiliation{F\'{\i}sica Te\'{o}rica, Universidad de Sevilla,
Apartado de Correos 1065, E-41080 Sevilla, Spain}
\date{\today}
\begin{abstract}
A one dimensional lattice model is formulated to study tapping dynamics and the
long time steady distribution in granular media. The dynamics conserves the
number of particles in the system, and density changes are associated to the
creation and destruction of empty sites. The model is shown to be consistent
with Edwards thermodynamics theory of powders. The relationship with lattice
models in which the number of particles is not conserved is discussed.
\end{abstract}

\maketitle

\section{Introduction}

In the last years, a great deal of effort is being made trying to
understand the physical mechanisms leading to compaction in weakly
vibrated granular systems, and the properties of the steady state
eventually reached in the long time limit. This has been prompted
and stimulated by the seminal papers of the Chicago group
reporting experimental results of compaction
\cite{KFLJyN95,NKPJyN97,NKBJyN98}. Granular compaction consists in
the increase of the density, starting from an initial low-density
state, as a consequence of external excitations, usually vertical
shakes or taps. Every tap is followed by a free relaxation, so
that in the process the system goes through a series of blocked
configurations.

Starting from an ``ergodic hypothesis'' for powders, based on the extensive,
global, character of the dynamics induced by shaking, Edwards and coworkers
\cite{EyO89} have formulated a microscopic theory for the steady state of
vibrated granular media that is similar to conventional statistical mechanics.
Moreover, they assume that the steady state is fully characterized by the
volume of the system, which plays then a role analogous to that of the energy
in  usual thermal systems. This provides the ``microcanonical'' description.
The associated ``canonical'' probability distribution is obtained by maximizing
the statistical entropy under the condition that the average volume is given.
Of course, the probability of a given configuration depends only of its volume.
The parameter conjugated of the volume, similar to the thermal temperature, was
named \textit{compactivity} by the authors in Ref. \cite{EyO89}.

Up to now, there is not a definite experimental test of the above thermodynamic
theory of powders. The measure of the compactivity, or the entropy, of a
granular system seems a rather difficult task not only in real experiments but
also in realistic models, although some procedures have been proposed. They are
based on the determination of the average volume and its fluctuations as a
function of the control parameter of the shaking, e.g. the vibration intensity
\cite{KFLJyN95,CyN01}. From these two functions, the compactivity can be
obtained, in principle, by integration, although this program is hard to carry
out in practice due to the uncertainty of the measurements. Another alternative
way, this one based on the generalization of the Einstein relation between
diffusivity and mobility, has been recently discussed and analyzed in a system
of inelastic hard spheres by means of molecular dynamics simulations
\cite{MyK02}.

On the other hand, the validity of Edwards theory has been studied in the
context of several simple models, with different underlying physical
mechanisms. It has been found that  the results for Tetris and spin-glass
models \cite{BKLyS00,FNyC02,DyL03} are consistent with the theory. In these
systems, the number of particles is fixed but most of the results are
numerical, due to the complexity of the models used. One dimensional Ising
models, with and without kinetic constrains, have also been considered
\cite{BPyS00,LyD01,BFyS02,SGyL02,PyB02}, because they are simple enough as to
allow a detailed analytical study in many cases. For weak tapping, agreement
with Edwards's theory was found again, although discrepancies show up in the
limit of strong tapping. Quite interestingly, all the Ising-like models in
Refs.\ \cite{BPyS00,LyD01,BFyS02,SGyL02,PyB02} have been formulated as open
systems. The number of particles does not remain constant, but it changes along
the compaction process, as a consequence of adsorption-desorption events from a
theoretical particle reservoir in contact with the system. Instead, it is the
volume what is kept fixed, leading in this way to the variation of the density.
Then, although it is true that the steady distribution of these models can be
considered as a ``grand canonical'' ensemble generalization of the theory, it
is also clear that it is not characterized by the compactivity (temperature)
but by another parameter playing the role of the chemical potential. This
difference is evidently relevant when trying to relate any of them with the
characteristics of the vibration process, e.g. its intensity. Beyond that, the
distinction might become conceptually crucial when dealing with granular
mixtures and segregation phenomena. In that case, each of the different species
is going to have its own analogous of the chemical potential parameter
\cite{PyB03}. Whether or not it is also needed to consider different
compactivities (temperatures), as it as been suggested recently \cite{NFyC02},
is a different question.

The aim of this paper is to present a closed, constant number of particles,
one-dimensional lattice model for compaction. Again the model is simple enough
as to be analytically tractable. During the tapping process, particles diffuse
and also empty sites (holes) are created and destroyed in the system, according
with well defined rules. The latter are chosen to mimic, in a crude way, what
happens in real compaction experiments. More precisely, the model tries to
represent a vertical section of a vibrated two-dimensional system. In a shake,
the length of the section can increase because empty regions (holes) are
created between particles. These regions can be used for the particles to
diffuse. Afterwards, once the shake has ended, the system tries to compact due
to the action of gravity. This is accomplished in the model by means of the
elimination of holes. But, because of the geometrical constrains following from
excluded volume effects of the hard particles in the neighbouring vertical
sections, not all the holes can be destroyed in the free evolution. Only large
enough empty regions can be reduced. As a consequence of the combination of a
tap and the next free relaxation, the length can increase in some regions of
the system and decrease in others. The net balance determines the global
behaviour of the system in the compaction process.

The plan of the paper is as follows. In Sec. \ref{s2}, the model is formulated
at the mesoscopic level of description  by means of a master equation for the
transition probability. This equation is exactly solved for the steady
distribution in Sec. \ref{s3}, and the associated macroscopic description is
discussed in Sec.\ \ref{s4}, where it is shown to be in agreement with Edwards'
thermodynamic description. The compactivity is identified in terms of the
parameters characterizing the mesoscopic dynamics of the model. Also, the
distribution of domains is there derived. Section \ref{s5} contains a detailed
discussion of the relationship between closed and open models, and between the
compactivity and fugacity parameters. The paper ends with a short summary and
some additional discussions.

\section{The model}
\label{s2}

We consider a one-dimensional lattice having $N+1$ particles. The number of
sites in the lattice is variable and changes in time accordingly with the rules
to be specified below. Those sites that are not occupied by a particle are said
to be empty or, equivalently, being a hole.

The dynamics of the system is defined trying to mimic tapping
experiments for the study of compaction in granular media
\cite{KFLJyN95,NKPJyN97,NKBJyN98}. These experiments typically
involve two different series of processes of quite different
nature. The system is submitted to taps or pulses separated by
time intervals for which the system is allowed to relax freely,
until being trapped in a metastable configuration. Therefore, each
tap starts in the metastable configuration reached in the previous
free relaxation. The taps are characterized by their duration and
their amplitude.

Physically, the effect of the taps is to decrease the local density in some
regions of the system, moving grains from their metastable positions, and
allowing a posterior reordering in the free relaxation. We will specify first
the dynamics during the relaxation processes, since it leads to identify the
possible metastable, or blocked, configurations of the model. It will be
assumed that in the free relaxation the system tries to reduce its length by
eliminating some of the empty sites of the lattice. More precisely, whenever
there is a group of nearest neighbour holes, all except one are eliminated.
These are the only processes taking place in the free relaxation, and have
probability one. Therefore, the number of particles is conserved in the
relaxation, but the length of the system, measured by the total number of
sites, is in general reduced.

As a consequence of the above dynamics for relaxation, the metastable
configurations of the model are characterized by having all the holes
surrounded by two particles, i.e. the holes are isolated. In order to displace
the system from one of these states, it has to be externally perturbed, for
instance by means of a tap. To complete the description of the dynamics of our
model in a compaction experiment, the possible transitions taking place during
a tap and starting at a mestastable configuration, have to be identified and
their probabilities specified. Two kinds of elementary processes will be
considered. Each of them will be discussed separately in the following.

\begin{figure}
\includegraphics[scale=0.5]{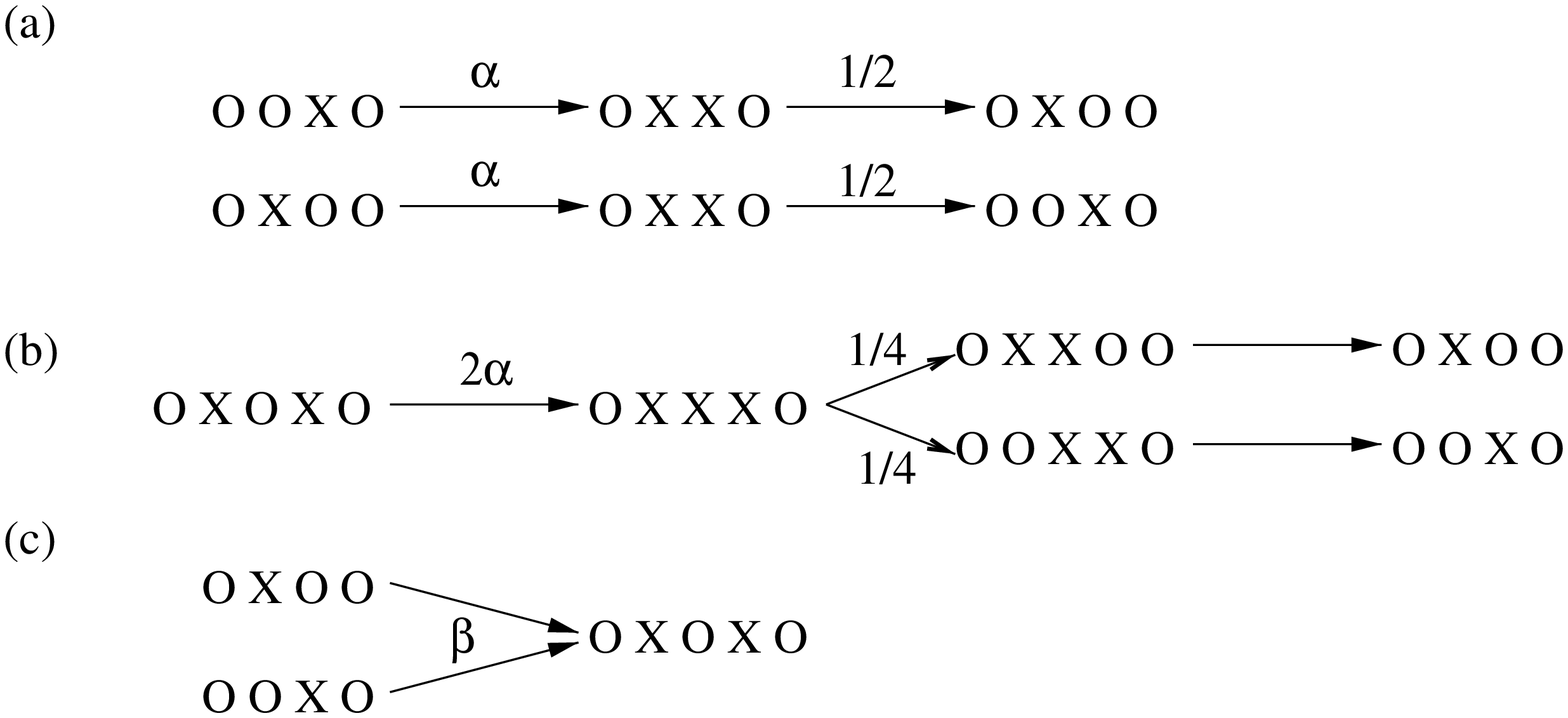}%
\caption{ \label{fig1} Elementary rearrangements of the system in a single tap
and the following free relaxation, in the weak vibration limit. The
trajectories leading to a final state identical to the initial one are not
shown.}
\end{figure}

Firstly, a particle can be transiently desorbed from the lattice and
posteriorly adsorbed in an empty site in the neighbourhood of its previous
position. This process is restricted by the following rule. A particle can be
desorbed from a site during the tap only if at least one of its nearest
neighbour sites is empty. More precisely, the probability for these events is
proportional to the number of nearest neighbour holes of the particle being
desorbed. This restriction tries to naively model the short range constraints
making difficult structural rearrangements in granular materials. Then, during
a tap, the probability of desorption of a particle having only one nearest
neighbour hole is $\alpha$, while it is $2 \alpha$ if it is surrounded by two
holes. Afterwards, the particle is reabsorbed either in its own original site
or in any of the nearest neighbour holes, with a probability that is
proportional to the number of holes next to the site considered. In Fig.
\ref{fig1}, cases $(a)$ and $(b)$ involve processes starting with the
transitory desorption of a particle. Particles and holes are represented by
circles and crosses, respectively. In the case referred to as $(b)$ in the
figure, the elimination of a hole happening in the next free evolution has been
also indicated. It is seen that the net result of the series of events taking
place during the tap and the free relaxation is, in this case, the destruction
of a lattice site, with the consequent decrease of the lattice length.

During the tap, the creation of an empty site or hole is also possible, but
only between two nearest neighbour particles located at one of the ends of a
domain of at least two particles. The probability of the corresponding
elementary events, referred to as case $(c)$ in Fig. \ref{fig1}, is $\beta$.
Note that these processes of hole creation are just the inverse of those
producing the destruction of a hole.

Furthermore, it will be assumed that only one transition takes place at the
most in every region on the system during each tap, i.e. no site is involved
into two different processes in the same perturbation of the system.
Physically, this hypothesis implies to consider the limit of weak and short
taps \cite{BPyS00,PyB02}. In summary, the dynamics of the model in the shaking
process is defined by the effective transitions given in Table \ref{table1},
describing the combined events associated to a tap and the next free
relaxation. The transitions only affect the clusters shown, and their
probabilities are independent of the configuration of the remainder of the
system.

To formulate the model in a more mathematical language, and also to
characterize the metastable configurations, it is convenient to define a set of
variables ${\bf n} \equiv \{n_{1},\cdots,n_{N} \}$. The variable $n_{i}$ takes
the value unity if there is a hole next to the right of particle $i$, while it
vanishes if there is no hole, i.e. if particles $i$ and $i+1$ are in nearest
neighbour sites. By definition, it is assumed that there is no hole to the left
of particle $1$ nor to the right of particle $N+1$. Both particles define the
boundaries of the system. It is easily realized that this property is preserved
by the dynamics of the system under tapping, as defined above. Then, we have
established a one to one relationship between a set of $N$ variables taking
values $0$ and $1$ and the metastable configurations of the model.

\begin{table}
\caption{\label{table1} Transition probabilities for the elementary
rearrangements taking place in a single tap, in the weak vibration regime.} \
\begin{ruledtabular}
\begin{tabular}{ccc}
Initial state & Final state & Probability \\
OOXO & OXOO & $\alpha/2$ \\
OXOO & OOXO & $\alpha/2$ \\
OXOXO & OXOO & $\alpha/2$ \\
OXOXO & OOXO & $\alpha/2$ \\
OXOO & OXOXO & $\beta$ \\
OOXO & OXOXO & $\beta$ \\
\end{tabular}
\end{ruledtabular}
\end{table}

The transition probabilities in Table \ref{table1} can be
expressed in terms of the $n_{i}$ variables. Denoting $R_{i}{\bf
n} \equiv \{ n_{1}, \cdots,R_{i}n_{i},\cdots,n_{N} \}$, with
$R_{i}n_{i}=1-n_{i}$, the probability $W({\bf n^{\prime}}|{\bf
n})$ of the several events going from configuration ${\bf n}$ to
configuration ${\bf n}^{\prime}$ in the effective dynamics
describing shaking process are:
\begin{equation}
\label{2.1} W(R_{i}R_{i+1} {\bf n}|{\bf n})= \frac{\alpha}{2}
\left[ n_{i}(1-n_{i+1})+(1-n_{i})n_{i+1} \right],
\end{equation}
\begin{equation}
\label{2.2} W(R_{i}{\bf n} |{\bf n})=\frac{\alpha}{2}
(n_{i-1}n_{i}+n_{i}n_{i+1})+\beta \left[ n_{i-1}(1-n_i)+(1-n_i)n_{i+1} \right]
\, .
\end{equation}
Equation (\ref{2.1}) corresponds to diffusive events, while the first and
second terms on the rhs of Eq.\ (\ref{2.2}) correspond to the destruction and
creation of holes, respectively. The Markov process defined by these transition
probabilities is irreducible, i.e. all the metastable configurations of the
lattice are connected by a chain of transitions with nonzero probability. To
verify this property, we begin by noting that any metastable configuration can
be connected with the configuration characterized by having just a hole located
next to the right of a given fixed particle. This is because holes can be moved
through particles by means of diffusive events, so that two consecutive holes
can always be located to both sides of the same particle. Afterwards, one of
the holes can be eliminated by a type $(b)$ event of Fig.\ \ref{fig1}. This
procedure can be repeated until there is only a hole in the lattice, that can
then be diffused to the desired site. This proves the above statement. But,
since each effective transition have its inverse also with nonzero probability,
the above paths can also be reversed, concluding that all the metastable
configurations are connected. The irreducibility property of the Markov process
implies that there is a unique steady probability distribution for the process
\cite{vK92}. This distribution will be explicitly obtained in the next Section.

\begin{figure}
\includegraphics[scale=0.5,angle=-90]{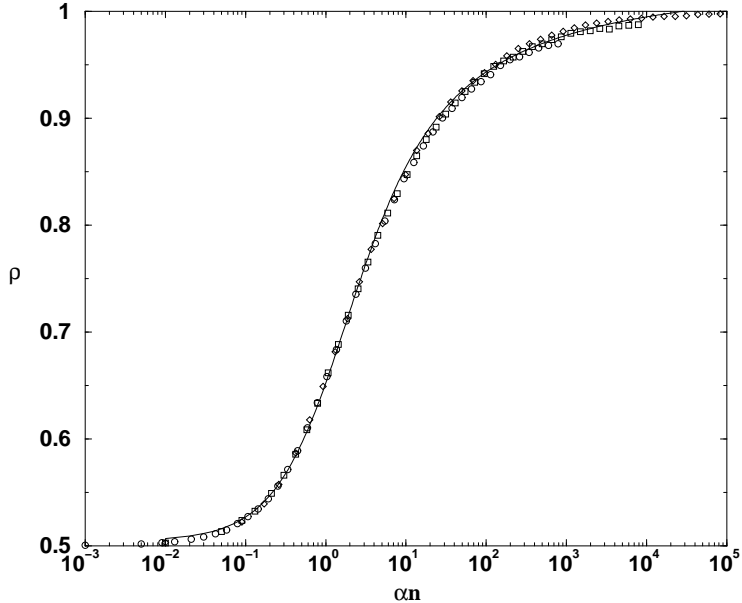}
\caption{\label{fig2} Evolution of the density of particles, as a function of
the scaled time defined in the text. The curves correspond to the pairs of
values $\alpha=10^{-3}$, $\beta=10^{-5}$ (circles), $\alpha=10^{-2}$,
$\beta=5\times 10^{-5}$ (squares) and $\alpha=10^{-2}$, $\beta=5\times 10^{-6}$
(diamonds), while the solid line is the best inverse logarithmic fit, Eq.
(\ref{2.3}), with the parameters given in the text.}
\end{figure}

In Fig.\ \ref{fig2} the relaxation of the particle density is shown as a
function of the ``scaled time'' $\tau=\alpha n$, where $n$ is the number of
taps before measuring the density,  for different values of $\alpha$ and
$\beta$. The initial state for all the curves was the least dense metastable
configuration, $\rho=0.5$, in which there is a hole between every two
particles. In all the reported cases $\beta\ll\alpha$, so that processes
decreasing the density of particles are only relevant when the system is near
the most compact state, $\rho=1$.  Moreover, as $\beta\ll\alpha\ll1$, there is
a universal behaviour up to a very large number of taps $t={\cal
O}(\alpha^{-1})$, i.e., $\alpha n={\cal O}(1)$. For longer times, when
processes decreasing the density become relevant, $n \gtrsim {\cal
O}(\beta^{-1})$, the system approaches a steady state characterized by the
ratio $\beta/\alpha$. The observed universal scaled curve is very well
described by the four-parameter empirical law
\begin{equation}
\label{2.3}
    \rho(t)=\rho_\infty-\frac{\delta\rho_\infty}{1+B\ln\left(1+\frac{\tau}{\tau_c}
     \right)} \, ,
\end{equation}
with $\rho_\infty=1.04$, $\delta\rho_\infty=0.54$, $B=1.17$ and $\tau_c=2.63$.
As it is the case with the experimental data \cite{KFLJyN95} and also with
numerical results from other simple models \cite{BPyS00,He00}, the logarithmic
fit is not expected to give the correct asymptotic density of particles. In
fact, in our case it is $\rho_\infty>1$, which is clearly unphysical. A similar
behaviour of $\rho_\infty$ was found in Ref. \cite{BPyS00}. Also, values of
$\rho_\infty$ larger than the random close packing limit have been reported
from the fit of experimental data \cite{KFLJyN95}.

\section{The steady distribution}
\label{s3}

To find the steady distribution of the Markov process describing
the effective dynamics of the model, we are going to suppose it
verifies detailed balance. Of course, this has to be justified a
posteriori by showing that such a distribution exists. Therefore,
we look for a time-independent distribution $p^{(s)}({\bf n})$
having the property
\begin{equation}
\label{3.1} W({\bf n}^{\prime}|{\bf n})p^{(s)}({\bf n})= W({\bf n}|{\bf
n}^{\prime})p^{(s)}({\bf n}^{\prime}) \, ,
\end{equation}
for all configurations $\bf n$ and ${\bf n^\prime}$. A direct first consequence
of this equation is that all the metastable configurations with the same number
of holes have the same probability in the steady state. This follows from the
fact that they are connected by diffusive events and diffusion is isotropic in
the effective dynamics, as seen in Table \ref{table1}. Therefore, the
distribution function verifying Eq.\ (\ref{3.1}) can only depend on the number
of holes,
\begin{equation}
\label{3.2} N_{H}=\sum_{i}^{N}n_{i},
\end{equation}
in the configuration, but not on their spatial distribution. So,
we can write
\begin{equation}
\label{3.3} p_{N_{H}}^{(s)}({\bf n})=\frac{f(N_{H})}{Z},
\end{equation}
where the number $N_{H}$ of holes in the configuration ${\bf n}$
has been made explicit in the notation, $f(N_{H})$ is a function
to be determined, and $Z$ a normalization constant,
\begin{equation}
\label{3.4} Z=\sum_{\bf n} f(N_{H}).
\end{equation}
Still remains to be analyzed if Eq.\ (\ref{3.3}) can be made
compatible with Eq.\ (\ref{3.1}), when particularized for
effective events increasing (and decreasing) the number of holes.
The latter reads
\begin{equation}
\label{3.5} \beta p_{N_{H}}^{(s)}({\bf n})=\frac{\alpha}{2}\,
p_{N_{H}+1}^{(s)}({\bf n}^{\prime}).
\end{equation}
Here ${\bf n}^{\prime}$ is a configuration differing from ${\bf
n}$ by the creation of a hole. Use of Eq.\ (\ref{3.3}) gives
\begin{equation}
\label{3.6} \frac{f(N_{H}+1)}{f(N_{H})}=\frac{2\beta}{\alpha}
\end{equation}
and, by iteration,
\begin{equation}
\label{3.7} f(N_{H})=C \left( \frac{2\beta}{\alpha}
\right)^{N_{H}},
\end{equation}
for $N_{H} \geq 1$, $C$ being an arbitrary constant that will be
taken equal to unity. In this way, we have proven that the system
has the property of detailed balance and that its steady
distribution is given by
\begin{equation}
\label{3.8} p_{N_{H}}^{(s)}({\bf n})=\frac{\gamma^{-N_{H}}}{Z},
\end{equation}
\begin{equation}
\label{3.9} Z=\sum_{\bf n} \gamma^{-N_{H}} =\sum_{N_{H}=1}^{N}
\Omega_{N}(N_{H}) \gamma^{-N_{H}},
\end{equation}
where $\gamma = \alpha/2\beta$ and $\Omega_{N} (N_{H})$ is the number of
metastable configurations of the lattice having $N_{H}$ holes and, of course,
$N+1$ particles. It is worth remarking that no approximation has been done in
order to derive the steady distribution, Eq. (\ref{3.8}), i.e, it is valid for
any value of $\gamma$. The steady average number of holes and its dispersion
can be evaluated from $Z$ by
\begin{equation}
\label{3.10} \langle N_{H}\rangle_{s} \equiv \sum_{\bf n} N_{H}
p^{(s)}_{N_{H}}({\bf n}) =-\frac{\partial \ln Z}{\partial \ln \gamma}\, ,
\end{equation}
\begin{equation}
\label{3.11} \langle\left(\Delta N_{H}\right)^{2}\rangle_{s} \equiv \langle
N_{H}^{2}\rangle_{s}-\langle N_{H}\rangle_{s}^{2} =\frac{\partial^{2} \ln
Z}{\partial (\ln \gamma)^{2}}=-\frac{\partial \langle N_{H}
\rangle_{s}}{\partial \ln \gamma}\, .
\end{equation}

A simple combinatorial argument gives
\begin{equation}
\label{3.12} \Omega_{N}(N_{H})= \frac{N!}{N_{H}!(N-N_{H})!}\, ,
\end{equation}
and substitution of this expression into Eq.\ (\ref{3.9}) yields
\begin{equation}
\label{3.13} Z= \left( 1+\frac{1}{\gamma} \right)^{N}-1.
\end{equation}
Therefore, it follows by application of Eq.\ (\ref{3.10}) that, in
the limit of large $N$,
\begin{equation}
\label{3.14} \langle N_{H}\rangle_{s}=\frac{N}{1+\gamma}\, .
\end{equation}
The right hand side of Eq.\ (\ref{3.14}) is a monotonic decreasing function of
$\gamma$, for fixed number of particles $N$, i.e., the length of the system
decreases as $\gamma$ increases. Therefore, $\gamma^{-1}$ plays in the model a
role similar to the vibration intensity in real granular experiments of
compaction. It follows that, in the physical image depicted by the present
model, the probability of diffusion proceses $\alpha$ is expected to grow
faster with the vibration intensity than the probability of creation of holes
$\beta$, al least in the weak tapping limit.

\begin{figure}
\includegraphics[scale=0.5,angle=-90]{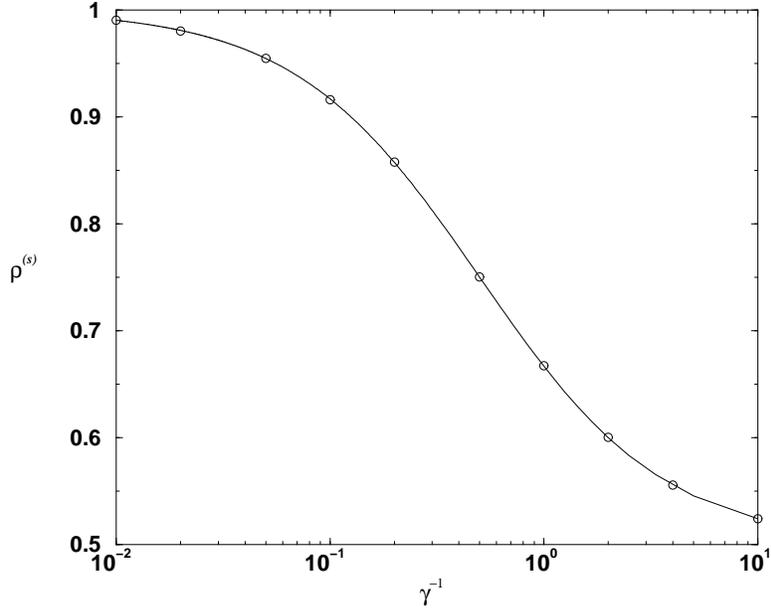}
\caption{\label{fig3} Comparison between the numerical values of the steady
density of particles and the theoretical prediction, given by Eq.\
(\ref{3.15b}).}
\end{figure}

The length (volume) of a configuration is given by
\begin{equation}
\label{3.15a}
    L=N_H+N \, ,
\end{equation}
and the average particle density is
\begin{equation}
\label{3.15b}
    \rho^{(s)}=\frac{N}{\langle L\rangle_s}=
    \frac{N}{\langle N_H\rangle_s+N}=
    \frac{1+\gamma}{2+\gamma} \, .
\end{equation}
In Fig.\ \ref{fig3}, the numerical values for the steady density of particles,
obtained by Monte Carlo simulation of the model, are compared with the
theoretical prediction, Eq.\ (\ref{3.15b}), and an excellent agreements is
found. The specific length per particle, in site units, is the inverse of the
particle density,
\begin{equation}
\label{3.15} \langle l\rangle_{s} \equiv \frac{ N+\langle
N_{H}\rangle_{s}}{N}=\frac{2+\gamma}{1+\gamma}\, .
\end{equation}
Its dispersion is obtained from Eqs. (\ref{3.11}) and
({\ref{3.13}),
\begin{equation}
\label{3.16} N \langle(\Delta l)^{2}\rangle_{s}=(2-\langle
l\rangle_{s})(\langle l\rangle_{s}-1),
\end{equation}
presenting a maximum for $\langle l\rangle_{s}=3/2$, i.e. when the average
number of holes is $N/2$ and the density of particles $\rho^{(s)}=2/3$, i.e.,
$\gamma=1$. The numerical evaluation of the length fluctuations is compared
with the theoretical prediction, as given by Eq.\ (\ref{3.16}), in Fig.\
\ref{fig4}. Again, a very good agreement is observed for the range of
``vibration intensities'' $\gamma^{-1}$ plotted. Outside this window of
vibration intensities, the length fluctuations are very small and, therefore,
rather hard to measure in the simulations.

\begin{figure}
\includegraphics[scale=0.5,angle=-90]{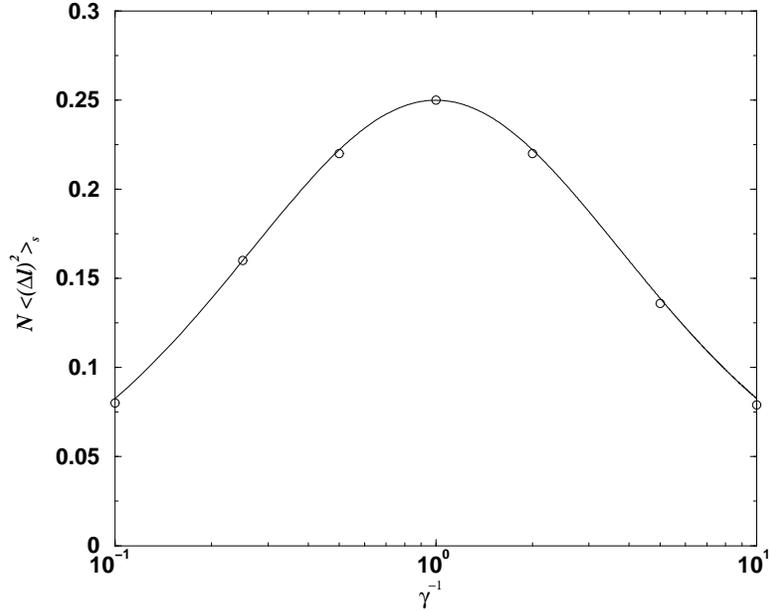}
\caption{\label{fig4} Comparison between the numerical evaluation of the length
fluctuations and the theoretical prediction, given by Eq.\ (\ref{3.16}). }
\end{figure}

\section{Thermodynamic description}
\label{s4}

Following Edwards and coworkers ideas \cite{EyO89}, the steady distribution
(\ref{3.8}) can be expressed in the ``canonical'' form
\begin{equation}
\label{3.17} p_{N_{H}}^{(s)}({\bf n})=\frac{e^{-N_{H}/X}}{Z},
\quad Z=\sum_{N_{H}} \Omega_{N}(N_{H}) e^{-N_{H}/X},
\end{equation}
where
\begin{equation}
\label{3.18}
 X= (\ln \gamma)^{-1}
\end{equation}
is the so-called compactivity. It is the conjugated thermodynamic parameter of
the volume in gently vibrated granular systems, in an analogous way as the
temperature is the conjugate of the energy in usual thermal systems. Note that,
in Eq.\ (\ref{3.17}), the ratio $N_{H}/X$ can be replaced in both the numerator
and the denominator by $L/X$, where $L$ is the length (volume) of the
configuration, as defined in Eq. (\ref{3.15a}). The structure of the above
steady distribution is consistent with the two main ingredients of Edwards'
theory, namely that the measure over metastable configurations is flat, and
that there is a unique parameter, the volume, characterizing the macroscopic
state of the system. Let us point out that, very recently, the theory has been
extended to include several macroscopic control parameters, in an effort to
explain the discrepancies observed in some models with strong tapping
\cite{BFyS02,Le02}, and also segregation patterns in binary models
\cite{NFyC02}. It is clear that such an extension does not apply to our model,
that is designed to describe compaction in one-component systems under weak
tapping. In the same context, the expression of the compactivity in Eq.
(\ref{3.18}) deserves some comments. Although $X$ can be formally negative, for
values $\gamma<1$, it is quite doubtful that this fact be physically relevant,
since this range of values of $\gamma$ corresponds to strong tapping, leading
to low stationary densities, namely with a average number of holes $\langle
N_{H}\rangle_{s}>N/2$. The possibility of negative values of the compactivity
has been also found in other simple models \cite{BPyS00,MyP97}, and it is
associated with the existence of a maximum length for the metastable
configurations.

In the limit $\gamma \rightarrow \infty$, i.e., asymptotically
weak tapping, the steady concentration of particles, $\rho^{(s)}$,
given by Eq.\ (\ref{3.15b}), can be approximated by
\begin{equation}
\label{3.19} \rho^{(s)}\simeq 1-\frac{1}{\gamma},
\end{equation}
and, using the definition in Eq.\ (\ref{3.18}),
\begin{equation}
\label{3.20} X^{-1}=- \ln (1-\rho^{(s)}).
\end{equation}
This relation between the compactivity and the steady density has been also
found in a model with facilitated dynamics having a variable number of
particles and fixed volume \cite{BPyS00}, and a similar behaviour has been
reported from the analysis of experimental data \cite{NKBJyN98}.

An ``entropy'' $S$ associated to the distribution $p^{(s)}$ can be
defined in the usual way,
\begin{equation}
\label{3.21} S=-\sum_{\bf n} p^{(s)}({\bf n}) \ln p^{(s)} ({\bf
n}),
\end{equation}
and use of Eq.\ (\ref{3.17}) gives
\begin{equation}
\label{3.22} S=\frac{\langle N_{H} \rangle_{s}}{X}+\ln Z.
\end{equation}
Taking into account Eq.\ (\ref{3.10}), it is easily verified that
\begin{equation}
\label{3.23} \frac{\partial S}{\partial \langle  L \rangle_{s}} =\frac{\partial
S}{\partial \langle N_{H} \rangle_{s}}=\frac{1}{X},
\end{equation}
consistently with the physical meaning of the compactivity as
discussed above. Given that the macroscopic state of the system is
characterized by a single parameter, it is possible to express the
entropy in terms of only the density of particles, or the
intensity parameter $\gamma$, or the compactivity. Then, for
instance, in the limit of large $N$ the entropy can be written as
\begin{equation}
\label{3.24} \frac{S}{N}= \frac{1}{X(1+e^{1/X})}+\ln
\frac{1+e^{1/X}}{e^{1/X}}.
\end{equation}

In addition to the global properties considered up to now, it is
also possible to obtain information about the domain structure of
the steady configurations. In particular, we are going to derive
here the probability distribution for the number of particles in a
domain. A domain of size $r$ is defined as a cluster of $r$
particles, i.e. two holes with $r$ particles in between. First, we
consider the probability $F_{r}^{(s)}$ of finding a local domain
of size $r$,
\begin{equation}
\label{3.25} F_{r}^{(s)}\equiv \langle n_{k}(1-n_{k+1})\cdots
(1-n_{k+r-1})n_{k+r} \rangle_{s},
\end{equation}
with $r \leq1$. Use of Eqs.\ (\ref{3.8}) and (\ref{3.12}) yields
\begin{equation}
\label{3.26} F_{r}^{(s)}=\frac{1}{Z} \sum_{N_{H}=0}^{N-r-1} \Omega_{N-r}(N_{H})
\gamma^{-2-N_{H}}=\gamma^{-2} \left( \frac{\gamma}{1+\gamma} \right)^{r+1},
\end{equation}
where the limit of large $N$ has been considered once again. Then, the
probability of a domain of size $r$, $P(r)$, is given by the conditional
probability of finding a cluster of $r$ consecutive particles plus a hole to
the right of a given hole, i.e.,
\begin{equation}
\label{3.27} P(r)=\frac{F_{r}^{(s)}}{\langle n_{k}
\rangle_{s}}=\frac{N}{\langle N_{H} \rangle_{s}}\,
F_{r}^{(s)}=\frac{\gamma^{r-1}}{(1+\gamma)^{r}}.
\end{equation}
This distribution is correctly normalized:
\begin{equation}
\label{3.28} \sum_{r=1}^{\infty} P(r) =1.
\end{equation}
It is instructive to express the distribution of domain sizes in terms of the
average length per particle, $\langle l \rangle_{s}$. This is easily
accomplished by means of Eq.\ (\ref{3.15}), obtaining
\begin{equation}
\label{3.29} P(r)=\left( 2-\langle l \rangle_{s} \right)^{r-1} \left( \langle l
\rangle_{s}-1 \right).
\end{equation}

\section{Relationship between closed and open models for compaction}
\label{s5}

In the previous Section, we have introduced the compactivity $X$ from the
canonical form of the steady probability distribution, Eq.\ (\ref{3.17}). In
the Edwards and coworkers formulation of the granular thermodynamic theory
\cite{EyO89}, the compactivity  was defined by
\begin{equation}
\label{5.1} X^{-1}=\left(\frac{\partial S}{\partial V}
\right)_{N},
\end{equation}
where the entropy $S$ is given by
\begin{equation}
\label{5.2} S=\ln \Omega_{N},
\end{equation}
$\Omega_{N}$ being the number of blocked configurations or, in the language
used in this paper, metastable states. In Eq.\ (\ref{5.1}), the number of
particles in the system is kept constant. The quantity $\Omega_{N}$ for the
model considered in this paper is given by Eq.\ (\ref{3.12}), and for $N \gg
1$, $N_{H} \gg 1$, Eq. (\ref{5.1}) leads to
\begin{equation}
\label{5.3} X^{-1}=\ln \frac{N-N_{H}}{N_{H}}.
\end{equation}
This is the ``microcanonical'' (constant volume) version of the ``canonical''
(constant compactivity) expressions (\ref{3.14}) and (\ref{3.18}). In fact,
combination of these two latter expressions gives
\begin{equation}
\label{5.4} X^{-1}=\ln \frac{N-\langle N_{H} \rangle_{s}}{\langle N_{H}
\rangle_{s}}.
\end{equation}
The expression equivalent to Eq.\ (\ref{5.2}) in the canonical
ensemble is Eq.\ (\ref{3.21}). It is evident that, in the limit of
large systems, it is consistent with the definition of $X$ in Eq.
(\ref{5.1}).

In several proposed models for compaction, lattices with a fixed number of
sites, i.e. fixed length, have been considered. The dynamics is defined
involving elementary processes associated with the adsorption and desorption of
particles, in such a way that the number of particles in the lattice changes
along the shaking experiment. This is the mechanism for which the density in
the system varies in time. In particular, several models leading to similar
kind of metastable configurations as in the model in this paper have been
discussed in detail \cite{BPyS00,LyD01,BFyS02,Le02,PyB02}. Then, aside from
details that are irrelevant for the following analysis, the number of blocked
configurations is given by Eq.\ (\ref{3.12}), that  we rewrite in the form
\begin{equation}
\label{5.5} \Omega_{ L}(N_{H})=\frac{( L-N_{H})!}{N_{H}! ( L-2N_{H})!},
\end{equation}
where the number of sites of the lattice $ L$ is now considered as fixed and $
L/2 \geq N_{H} \geq 1$. Moreover the steady distribution, in the weak tapping
limit was found to have the form
\begin{equation}
\label{5.6} p^{(s)\prime}({\bf
n})=\frac{\eta^{-N_{H}}}{Z^{\prime}},
\end{equation}
with
\begin{equation}
\label{5.7} Z^{\prime}=\sum_{N_{H}=1}^{ L/2} \Omega_{N}(N_{H}) \eta^{-N_{H}},
\end{equation}
and $\eta$ is a given parameter, depending on the specific model, and
characterizing the dynamical events in the system under shake. Then, from Eq.
(\ref{5.6}) a compactivity $X^{\prime}$ was identified as
\begin{equation}
\label{5.8} X^{\prime} = (\ln \eta )^{-1}.
\end{equation}
This definition is equivalent to
\begin{equation}
\label{5.9} X^{\prime -1}=\left( \frac{\ln \Omega_{ L}(N_{H})}{\partial N_{H}}
\right)_{ L} =-\left( \frac{\partial S}{\partial N} \right)_{ L},
\end{equation}
or, using Eq.\ (\ref{5.5}),
\begin{equation}
\label{5.10} X^{\prime -1}=\ln\, \frac{(N-N_{H})^{2}}{N_{H}N}.
\end{equation}
This expression differs from Eq.\ (\ref{5.3}) except in the limit of high
density $N_{H}/N \ll 1$, in which both reduce to $\ln (N/N_{H})$, But this only
indicates that the same density is obtained in this limit if $X=X^{\prime}$,
although it must be stressed that $X$ is associated to tapping processes at
constant number of particles, while $X^{\prime}$ describes processes at
constant volume. Equivalently, $X$ characterizes ensembles with fixed $N$, and
$X^{\prime}$ ensembles with fixed $L$. In this context, their physical nature
is rather different. The parameter $X$ is the compactivity introduced by
Edwards and, on the other hand, $\eta^{-1}$, related with $X^{\prime}$ by Eq.\
(\ref{5.8}), plays the role of a fugacity for the particles. In terms of the
entropy, $X$ and $X^{\prime}$ are related by
\begin{equation}
\label{5.11} X^{-1}=X^{\prime -1} + \left( \frac{\partial
S}{\partial N} \right )_{N_{H}}.
\end{equation}

\section{Conclusion}

In this paper, a one-dimensional model for compaction in granular media has
been presented. One of its main features, as compared with previous Ising-like
models, is that the time evolution under tapping conserves the number of
particles in the system, while it is the volume what changes in the compaction
process. This is in fact what happens in compaction experiments. Consequently,
the steady distribution is characterized by the compactivity instead of by a
generalized fugacity. The steady distribution function has been derived and the
compactivity identified in terms of the parameters defining the mesoscopic
dynamics of the model. It has been found that the results are consistent with
Edwards' thermodynamical theory of powders. Nevertheless, since the model is
formulated in the context of weak and short tapping, it is in fact quite
doubtful that the same conclusion were reached from a generalization to
stronger tapping processes. Let us point out that this would require to modify
the formulation of our model by including the possibility that a lattice region
would experiment several elementary excitations during the same tap.

The relationship between closed and open models, and between compactivity and
fugacity, has been discussed. At the mesoscopic level of description used in
this paper, the expression of one of them in terms of the transition rates can
not be inferred from the expression of the other. Nevertheless, it is true that
they correspond to different derivatives of the same entropy function, like in
usual thermal systems.

The model presented here can be easily generalized to mixtures of several kinds
of grains, then allowing the study of segregation phenomena. Also, it can be
useful to investigate the validity of the Edwards theory in this case, and
eventually its possible generalizations, for instance by extending the number
of parameters needed to characterize the steady state of the mixture, as it has
been proposed recently \cite{NFyC02,PyB03}.

\begin{acknowledgments}
We acknowledge support from the Ministerio de Ciencia y
Tecnolog\'{\i}a (Spain) through Grant No. BFM2002-00303 (partially
financed by FEDER funds) .
\end{acknowledgments}

\end{document}